\documentclass{article}

\usepackage{amsmath}
\usepackage{amssymb}
\usepackage[utf8]{inputenc}
\usepackage{hyperref}
\usepackage{graphicx}
\usepackage{listings}
\usepackage{xcolor}
\usepackage{xspace}
\usepackage{comment}
\usepackage{placeins}
\usepackage{subcaption}

\DeclareMathOperator{\f}{f}
\DeclareMathOperator{\h}{h}
\DeclareMathOperator{\cost}{cost}
\DeclareMathOperator{\parCost}{parCost}
\DeclareMathOperator{\serCost}{serCost}
\DeclareMathOperator{\same}{same}
\DeclareMathOperator{\opp}{opp}

\newtheorem{property}{Property}

\newcommand\etal{\emph{et~al.\@}\xspace}

\newcommand{\beginsupplement}{%
	\setcounter{table}{0}
	\renewcommand{\thetable}{S\arabic{table}}
	\setcounter{figure}{0}%
	\renewcommand{\thefigure}{S\arabic{figure}}
	\renewcommand{\theHtable}{Supplement.\thetable}
	\renewcommand{\theHfigure}{Supplement.\thefigure}
}

\title{Exact and heuristic algorithms for Cograph Editing}
\author{W. Timothy J. White, Marcus Ludwig and Sebastian Böcker}

\begin{document}
\maketitle

\abstract{We present a dynamic programming algorithm for optimally solving the \textsc{Cograph Editing} problem on an $n$-vertex graph that runs in $O(3^n n)$ time and uses $O(2^n)$ space.
In this problem, we are given a graph $G = (V, E)$ and the task is to find a smallest possible set $F \subseteq V \times V$ of vertex pairs such that $(V, E \bigtriangleup F)$ is a cograph (or $P_4$-free graph), where $\bigtriangleup$ represents the symmetric difference operator.
We also describe a technique for speeding up the performance of the 
algorithm in practice.
Additionally, we present a heuristic for solving the \textsc{Cograph Editing} 
problem which produces good results on small to medium datasets. In 
application it is much more important to find the ground truth, not some 
optimal solution. For the first time, we evaluate whether the cograph 
property is strict enough to recover the true graph from data to which noise has been added.
}

\section{Introduction}

A \emph{cograph}, or \emph{complement reducible graph}, is a simple undirected graph that can be built from isolated vertices using the operations of disjoint union and complement.
Specifically:

\begin{enumerate}
    \item A single-vertex graph is a cograph.
    \item The disjoint union of two cographs is a cograph.
    \item The complement of a cograph is a cograph.
\end{enumerate}

There are several equivalent definitions of cographs \cite{corneil1981complement}, perhaps the simplest to state being that cographs are exactly the graphs that contain no induced $P_4$ (path on 4 vertices).
As a subclass of perfect graphs, they enjoy advantageous algorithmic properties: many problems that are NP-hard on general graphs, such as \textsc{Clique} and \textsc{Chromatic Number}, become polynomial-time for cographs.

Cographs can be recognised in linear time \cite{corneil1985linear}.
A more difficult problem arises when we ask for the minimum number of
``edge editing'' operations required to transform a given graph into a
cograph.
Three problem variants can be distinguished: If we may only insert edges, we 
have the \textsc{Cograph Completion} problem; if we may only delete edges, 
we have the \textsc{Cograph Deletion} problem; if we may both insert and 
delete edges, we have the \textsc{Cograph Editing} problem.
When framed as decision problems, in which the task is to determine whether
such a transformation can be achieved using at most a given number $k$ of
operations, all three problem variants are NP-complete
\cite{el-mallah1988complexity,liu2012complexity}.
(Note that the edge completion and deletion problems can be trivially 
transformed into each other by taking complements.)
A general result of Cai, when combined with linear-time recognition of cographs, directly gives an $O(6^kn)$ fixed-parameter tractable (FPT) algorithm \cite{cai1996fixed} for \textsc{Cograph Editing}; more recently, a $O(4.612^k + n^{4.5})$ FPT algorithm \cite{liu2012complexity} has been described.

Concerning applications, we focus in particular on a recently developed approach for inferring phylogenetic trees from gene orthology data that involves solving the \textsc{Cograph Editing} problem \cite{hellmuth2015phylogenomics}.
Briefly, in this setting we may represent genes as vertices in a graph, with pairs of vertices linked by an edge whenever they are deemed to have arisen through a speciation (as opposed to gene duplication) event.
In a perfect world, this graph would be a cograph, and its cotree (see below) would correspond to the gene tree, which can be combined with gene trees inferred from other gene families to infer a species tree.
In the real world, measurement errors---false positive and false negative inferences of orthology---frequently cause the inferred orthology graph not to be a cograph, and in this case it is reasonable to ask for the smallest number $k^*$ of edge edits that would transform it into one.

For practical instances arising from orthology-based phylogenetic analysis, it is often the case that $k^* > n$ or even $k^* = \Omega(m)$, limiting the effectiveness of FPT approaches parameterised by the number of edits and motivating the development of ``traditional'' exponential-time algorithms---that is, algorithms that require time exponential in the number of vertices $n$.
We first give a straightforward dynamic programming algorithm that solves the more general edge-weighted versions of each of the three problem variants in $O(3^n n)$ time and $O(2^n)$ space, and which additionally offers simple implementation and predictable running time and memory usage.
We then describe modifications that are likely to significantly improve running time in practice, without sacrificing optimality (though also without improving the worst-case bound).
In addition, we describe and evaluate a heuristic solving the 
\textsc{Cograph Editing} problem based on an algorithm by Lokshtanov \etal 
~\cite{lokshtanov10characterizing}.

\subsection{Definitions}

Every cograph $G = (V, E)$ determines a unique vertex-labelled tree $T_G = (U, D, \h : U \to \{0, 1\})$ called the \emph{cotree} of $G$, which encodes the sequence of basic operations needed to build $G$ from individual vertices.
The vertices of $T_G$ correspond to induced subgraphs of $G$: leaves in $T_G$ correspond to individual vertices of $G$, and internal vertices to the subgraphs produced by combining the child subgraphs in one of two ways, according to whether the vertex is labelled 0 or 1 by $\h$.
0-vertices specify \emph{parallel} combinations, which combine the subgraphs represented by the child vertices into a single graph via disjoint union, while 1-vertices specify \emph{serial} combinations, which combine these subgraphs into a single graph by adding all possible edges between vertices coming from different children (or equivalently, by complementing, forming the disjoint union, and then complementing again).
The root is labelled 1, and every path from the root alternates between 0-vertices and 1-vertices.

Given a cotree $T_G$, a postorder traversal that begins with a distinct single-vertex graph at each leaf and then applies the series or parallel combination operations specified at the internal nodes will culminate, at the root node, in the corresponding cograph $G$.

\section{An $O(3^n n)$-time and $O(2^n)$-space algorithm for weighted 
cograph editing, completion, and deletion problems}

We describe here an algorithm for the weighted version of the \textsc{Cograph Editing} problem.
The deletion and completion problem variants are dealt with using simple 
modifications to the base algorithm, described later.
Unweighted variants can of course be obtained by setting all edge weights to 1.

Given an undirected graph $G = (V, E)$ with $V = \{ v_1, \dots, v_n \}$ and vertex-pair weights given by $w : V \times V \to \mathbb R^{\ge 0}$, with the interpretation that $w(u, v)$ is the cost of deleting the edge $(u, v)$ when $(u, v) \in E$ and the cost of inserting it otherwise, we seek a minimum-weight edge modification set $F \subseteq V \times V$ such that $(V, E \bigtriangleup F)$ is a cograph, where $\bigtriangleup$ is the symmetric difference operator.

We compute the minimum cost of transforming every subset of vertices into a cograph using dynamic programming.
The algorithm hinges on the following property of cotrees \cite{Lerchs1972}:

\begin{property}\label{pro:lca}
In a cograph $G$, two vertices $u$ and $v$ are linked by an edge if and only if their lowest common ancestor in the cotree $T_G$ of $G$ is a series node.
\end{property}

For any subset $X$ of vertices in $V$, let $v_X$ denote the vertex with maximum index in $X$.  We can compute the minimum cost $\f(X)$ of editing the induced subgraph $G[X]$ to a cograph using:

\begin{equation}
\label{eqn:dp}
\f(X) =
\begin{cases}
0, & \text{if}\ |X| < 4\\
\min_{Y \subsetneq X, v_X \in Y} (\f(Y) + \f(X \setminus Y) + \cost(Y, X \setminus Y)), & \text{otherwise}
\end{cases}
\end{equation}
\begin{equation}
\cost(A, B) = \min \{ \parCost(Y, X \setminus Y), \serCost(Y, X \setminus Y) \}
\end{equation}
\begin{equation}
\parCost(A, B) = \sum_{\{(u, v) \in E : u \in A, v \in B\}} w(u, v)       \quad \text{(these edges need to be deleted)}
\end{equation}
\begin{equation}
\serCost(A, B) = \sum_{\{(u, v) \notin E : u \in A, v \in B\}} w(u, v)        \quad \text{(these edges need to be inserted)}
\end{equation}

Because each invocation of $\f$ has a strictly smaller set of vertices as input, it suffices to compute solutions to subproblems in increasing order of subset size.
To instead solve the edge deletion (respectively, insertion) problem, replace $\serCost$ (respectively, $\parCost$) with a function that is zero when the original function is zero, and infinity otherwise.

The $3^n$ factor in the time complexity arises from there being at most one argument to the outer $\min$ for every way of partitioning $V$ into 3 parts $(V \setminus X, X \setminus Y, Y)$.
If we enumerate bipartitions $(Y \mid X \setminus Y)$ in Gray code order, then straightforward algorithms for computing $\parCost$ and $\serCost$ incrementally may be used, resulting in the additional factor of $n$.

An optimal solution can be found by back-tracing the dynamic programming matrix as usual.
It is possible to extract every optimal solution this way, but producing each of them exactly once requires a slight reformulation whereby we include the root node type (series or parallel) in the dynamic programming state, which doubles the memory requirement.

Although the above is a ``subset convolution''-style dynamic program, the possibility of achieving $O^*(2^n)$ time by applying the Möbius transform approach of Björklund et al. \cite{bjorklund2007fourier} appears to be complicated by the third term in the summation.

\section{Reducing the number of partitions considered}

Given any subset $X$ of vertices, the dynamic programming algorithm above enumerates every possible bipartition to find a best one, and thereby compute $\f(X)$.
For many graphs encountered in practice, this will be overkill: The vast majority of bipartitions tried will be very bad, suggesting that there could be a way to avoid trying many of them without sacrificing optimality.
Instead of enumerating all bipartitions of $X$, we propose to use a search tree to gradually refine a \emph{set} of bipartitions defined by a series of weaker constraints, avoiding entire sets of bipartitions that can be proven to lead to suboptimal solutions.
Here we describe a branch and bound algorithm, running ``inside'' the dynamic program, that uses this strategy to find an optimal bipartition of a given vertex subset $X$.
We however note that, since $O(2^n)$ space is already needed by the ``core'' dynamic programming algorithm, and since a full enumeration of all bipartitions of $X$ would require only asymptotically the same amount of space, an $A^*$ algorithm is likely feasible.
The overall strategy is somewhat inspired by the Karmarkar-Karp heuristic \cite{KarmarkarKarp1982} for number partitioning, which achieves good empirical performance on this related problem by deferring as far as possible the question of exactly which part in the partition to assign an element to.

The basic idea is to maintain, in every subproblem $P$, a set of constraints $S_P$ of the form ``$A \subseteq X$ are all in the same part'', and another set of constraints $O_P$ of the form ``$A \subseteq X$ and $B \subseteq X$ are in different parts'' (clearly $A \cap B = \emptyset$).
We call a constraint of the former kind a \emph{$\same$-constraint} and denote it $\same(A)$; a constraint of the latter kind we call an \emph{$\opp$-constraint} and denote it $\opp(A | B)$.
Each subproblem (except the root; see below) has one extra bit of information $\lambda_P \in \{+,-\}$ that records whether it represents a series ($+$) or parallel ($-$) node in the cotree: we will see later that separating these two cases enables stronger lower bounds to be used.
A subproblem $P = (\lambda_P, C_P)$ thus represents the set of all bipartitions consistent with its \emph{constraint set} $C_P = S_P \cup O_P$ that introduce a series ($\lambda_P = +$) or parallel ($\lambda_P = -$) node in the cotree.
Note that any cotree may be represented as a binary tree with internal nodes labelled either $+$ or $-$ (though this representation is not unique).

The root subproblem $P_X$ for a vertex subset $X$ is special: it has no associated $\lambda$ value; rather it has has exactly two children $P_{X^+}$ and $P_{X^-}$ having opposite values of $\lambda$, with each containing only the trivial constraints $C_{P_{X^+}} = C_{P_{X^-}} = \{ \same(\{v\}) : v \in X \}$.
These two children (which may be thought of as the roots of entirely separate search trees) thus together represent the set of all possible configurations of $X$, where a configuration is a bipartition together with a choice of cotree node type (series or parallel).

\subsection{Generating subproblems}

Before discussing the general rule we use for generating subproblems, we first give a simplified example.
If there are two vertices $u$, $v$ in $X$ that have not yet been used in any $\opp$-constraint, we may create a new subproblem in which $u$ and $v$ are forced to be in the same part of the bipartition, as well as another new subproblem in which they are forced to be in opposite parts.
(Clearly every bipartition in the original subproblem belongs to exactly one of these two subproblems.)

\subsubsection{Structure of a general subproblem}

More generally, let $S^*_P$ be the set of all inclusion-maximal subsets of $X$ appearing in a $\same$-constraint in subproblem $P$ (i.e., $S^*_P = \{ Z \subseteq X : \same(Z) \in S_P \land (\nexists Z' \supsetneq Z \land \same(Z') \in S_P) \}$).
Then we may choose two distinct (necessarily disjoint and nonempty) subsets $A$ and $B$ from $S^*_P$ and form two new subproblems: one in which $\same(A \cup B)$ is added to the constraint set, and one in which $\opp(A | B)$ is added.
Each new subproblem inherits the $\lambda$ of the original.
As before, every bipartition in the original subproblem belongs to exactly one of these two subproblems.
When no such pair $(A, B)$ can be found, we halt this refinement process and enumerate all bipartitions consistent with the constraints, evaluating each as per Equation~\ref{eqn:dp}.
In this way, the constraint set $C_P$ of any subproblem $P$ can be represented as a directed forest, each component of which is a binary tree that may have either a $\same(\cdot)$ node or an $\opp(\cdot)$ node at the root and $\same(\cdot)$ nodes everywhere else.
The components containing only $\same(\cdot)$ nodes are exactly the members of $S^*_P$, so a standard union-find data structure \cite{tarjan1975efficiency} can be used to efficiently find a pair of vertex sets $A, B \in S^*_P$ eligible for generating a new pair of child subproblems.

Although it would be possible and perhaps fruitful to continue adding constraints of a more complicated form, for example a constraint $\opp(A|C)$ when the constraints $\opp(A|B)$ and $\opp(C|D)$ already exist in $C_P$, there are several reasons to avoid doing so.
First, allowing such constraints destroys the simple forest structure of constraints in $C_P$.
In the presence of such constraints, a new candidate constraint may be tautological or inconsistent; these cases can be detected (for example using a 2SAT algorithm), but doing so slows down the process of finding a new candidate constraint to add.
Second, any set of constraints containing one or more such complicated constraints is ``dominated'' in the sense that some set of simple constraints exists that implies the same set of bipartitions, meaning that no additional ``power'' is afforded by these constraints.

\subsection{Strengthening lower bounds}

The procedure described above is only useful in reducing the total number of bipartitions considered if the constraints added in subproblems are able to improve a lower bound on the cost of a solution.
The lower bound $L_P$ associated with any subproblem $P$ will have the form $L_P = \sum_{x \in S^*_P}{L_S(x)} + \sum_{\opp(A|B) \in O_P}{L_O(A, B)}$.
We now examine the two kinds of terms in this lower bound, and how they may be efficiently computed for a subproblem from its parent subproblem.

\subsubsection{Lower bounds $L_S(\cdot)$ from $\same$-constraints}

Whenever two $\same$-constraints $\same(A)$ and $\same(B)$ are combined into a single $\same$-constraint $\same(A \cup B)$, we may add $\f(A \cup B) - \f(A) - \f(B)$ to the lower bound.
This represents the cost of editing the entire vertex set $A \cup B$ into a cograph, offset by subtracting the costs already paid for editing each vertex subset $A$ and $B$ into cographs.
Note that any function computing a lower bound on these costs can be used in place of $\f$, provided that its value does not change between the time at which it is first added to the lower bound (at some subproblem), and later subtracted (at some deeper subproblem).
If the bottom-up strategy is followed for computing $\f$, then we always have these function values available exactly.

\subsubsection{Lower bounds $L_O(\cdot, \cdot)$ from $\opp$-constraints}

Whenever an $\opp$-constraint $\opp(A|B)$ is added to a parallel subproblem, then for each edge $(a, b) \in E$ with $a \in A$ and $b \in B$ we may add $w(a, b)$ to the lower bound.
This represents the cost of deleting these edges, which cannot exist if they are in different subtrees of a parallel cotree node by Property~\ref{pro:lca}.
Because the sets of vertices involved in the $\opp$-constraints of a given subproblem are all disjoint, no edge is ever counted twice.
The reasoning is identical for series subproblems, except that we consider all vertex pairs $(a, b) \notin E$: these edges must be added.

In fact it may be possible to strengthen this bound by considering vertices that belong neither to $A$ nor to $B$: For any triple of vertices $a \in A$, $b \in B$, $v \in X \setminus (A \cup B)$ such that neither $(v, a)$ nor $(v, b)$ is in $E$, we may in principle add $\min \{ w(v, a), w(v, b) \}$ to the lower bound, since $v$ cannot be in the same part of the bipartition as both $a$ and $b$ and so must, by Property~\ref{pro:lca}, have an edge to at least one of these vertices added.
However, doing so introduces the possibility of counting an edge multiple times.
Although this can be addressed, doing so appears to come at its own cost: For example, dividing by the maximum number of times that any edge is considered produces lower bound increases that are valid but likely weak; while partitioning vertices \emph{a priori} and then counting only bound increases from vertex triples in the same ``pristine'' part of the partition has the potential to produce stronger bounds, but entails significant extra complexity.

\subsection{Choosing a subproblem pair}

It remains to describe a way to choose disjoint vertex sets $A, B \in M_P$ to use for generating child subproblem pairs.
The strategy chosen is not important for correctness, but can have a dramatic effect on the practical performance of the algorithm.
Since both types of child subproblem are able to improve lower bounds, a sensible choice is to consider all $A, B \in M_P$ and choose the pair that maximises $\min \{ LB(S_P \cup O_P \cup \{ \same(A \cup B) \}), LB(S_P \cup O_P \cup \{ \opp(A | B) \}) \}$---that is, the pair $A, B$ that offers the best worst-case bound improvement.
Ties could be broken by $\max \{ LB(S_P \cup O_P \cup \{ \same(A \cup B) \}), LB(S_P \cup O_P \cup \{ \opp(A | B) \}) \}$.
Any remaining ties could be broken arbitrarily, or perhaps using more 
expensive approaches such as fixed-length lookahead.

\section{Heuristics for the \textsc{Cograph Editing} problem}

In the following we will assume unweighted graphs.
For graph $G = (V, E)$ we denote $V(G) = V$ and $E(G) = E$. Given vertices 
$V' \subseteq 
V$, $G[V']$ is the subgraph of $G$ induced by $V'$. All neighbors of $v$ in 
$G$ are denoted by $N_v(G)$. 

Lokshtanov \etal~\cite{lokshtanov10characterizing} developed an algorithm to 
find a minimal cograph completion $H = (V, E \cup F)$ of $G$ in $O 
(|V|+|E|+|F|)$ time.
A cograph completion is called minimal if the set of added edges $F$ is 
inclusion minimal, i.e., if there is no $F' \subsetneq F$ such that
$(V, E \cup F')$ is a cograph.
Let $G_x$ be a graph obtained from $G$ by adding a vertex $x$ and connecting 
it to some vertices already contained in $G$, and let $H$ be any minimal
cograph completion of $G$. 
Lokshtanov \etal showed that there exists a minimal cograph 
completion $H_x$ of $G_x$ such that $H_x[V(H)] = H$.

Using this observation they describe a way to compute a minimal cograph 
completion of $G$ in an iterative manner starting from an empty graph 
$H_0$.
In each iteration a graph $G_{i}$ is derived from $H_{i-1}$ by adding a new 
vertex $v_{i}$ from $G$. Vertex $v_{i}$ is connected to all its neighbors 
$N_{v_i}(G[V(G_i)])$ in $G_i$. Now a set of additional edges $F_{v_i}$ is 
computed such that $H_i = (V(G_i), E(G_i) \cup F_{v_i})$ is a minimal 
cograph completion of $G_i$. Finally, $H_n$ is a minimal cograph completion 
of $G$.

It is obvious that finding a minimal cograph completion gives an 
upper bound on the \textsc{Cograph Editing} problem.
To find a minimal cograph deletion of $G$ we can simply find a minimal 
cograph completion of its complement $\bar{G}$. 
This algorithm allows us to efficiently find minimal cograph deletions and 
completions. However, only adding or only deleting edges from $G$ is rather 
restrictive for finding a good heuristic solution for the \textsc{Cograph 
Editing} problem.
(Indeed, it is straightforward to construct instances for which
insertion-only or deletion-only strategies yield solutions that are
arbitrarily far from optimal.)
To allow a combination of edge insertions and deletions, in each iteration 
step we choose a vertex $v_i$ and compute the minimum set of edges $F_{v_i}(G_i)$ to make $H_i$ a 
minimal cograph completion of $G_i$. Furthermore, we compute 
$F_{v_i}(\bar{G_i})$ for its complement.
If $|F_{v_i}(G_i)| \leq |F_{v_i}(\bar{G_i})|$ we add edges to $H_i$. Else we 
remove edges from $H_i$ to preserve the cograph property.
In this way Lokshtanov's algorithm serves as a heuristic for the 
\textsc{Cograph Editing} problem, allowing us to add and remove edges from 
$G$.
Finding a set of edges $F_{v_i}$ such that $H_i$ is a 
minimum cograph completion of $G_i$ takes $O(|N_{v_i}(H_i)|+1)$ time.
Computing both, $F_{v_i}(G_i)$ and $F_{v_i}(\bar{G_i})$, needs $O(|V|)$ time.
Hence, allowing edge insertions and deletions in every step increases 
overall running time to $O(|V|^2)$.

To improve the heuristic there are multiple natural modifications which can 
be easily integrated into the Lokshtanov algorithm.
The resulting cograph clearly depends on the order in which vertices 
from $G$ are drawn. Although it is infeasible to test all possible 
orderings of vertices in $V$, it is nevertheless worthwhile to try more than
one.
In our simplest version of the heuristic, we draw random orderings from $V$ and compute a 
solution for each ordering. Going further, we can test multiple vertices in each 
iteration step and add the vertex $v_i$ to $H_i$ which needs the smallest 
number of 
edge modifications. In its most exhaustive version this leads to an 
algorithm which greedily takes in each step the best of all remaining 
vertices and adds it to $H_i$. This algorithm's running time increases by 
a factor of $O(|V|)$.
Another version of the heuristic may apply beam search, storing the best 
$k$ intermediate results in each step.

All modifications described so far restrict each iteration to performing
only insertions or only deletions. In order to search more broadly, when considering
how to compute $H_i$ from $G_i$, we may
test whether removing a single edge incident on $v_i$ in $G_i$ before 
inserting edges as usual results in fewer necessary edge edits overall.
A similar strategy can be applied to the complement graph. 
In this way we can insert and delete edges in a single step -- for the 
price 
of having to iterate over all of $v_i$'s neighbors. If we apply this 
strategy in each step to $G_i$ and its complement $\bar{G_i}$, the running 
time increases by a factor of $O(|V|)$.

It must be noted that by applying Lokshtanov's algorithm in the above 
manner, we lose any proven guarantees such as minimality.

\subsection{Results}

We evaluate five heuristic versions. All of them consider adding edges or 
deleting edges in each iteration. Unless stated 
otherwise we run each heuristic 100 times and take the cograph with lowest 
costs. The five versions are:
\begin{enumerate}
	\item \textit{standard}: Compute a cograph using random vertex insertion 
	order.
	\item \textit{modify}: When adding $v_i$, allow removing one vertex from 
	$v_i$'s neighborhood in $G_i$. Hence, multiple edges may be inserted and 
	a single 
	edge may be deleted in the same step (or vice versa when applied to 
	$\bar{G_i}$). 
	\item \textit{choose-multiple}: In each step, choose 10 random vertices,
	and add the one with the lowest modification costs.
	\item \textit{beam-search}: Maintain 10 candidate solutions.  In each 
	step, for each candidate solution choose 10 random vertices and try 
	adding each of them;
	keep the best 10 of the 100 resulting solutions.  Run the heuristic 10 
	times instead of 100. 
	\item \textit{choose-all}: In each step, consider all remaining vertices from $G$,
	and add the one with the lowest modification costs.
	Run the heuristic just once.
	%	\todo{running multiple times could still find different results by 
	%resolving ties.}
\end{enumerate}

We evaluate the heuristics on simulated data. As \textsc{Cograph Editing} is 
NP-complete it is computationally too expensive to identify the 
correct solution for reasonable size input graphs. We 
simulate cographs and afterwards randomly perturb edges. The true 
cograph serves as a proxy for the optimal cograph: It is a good bound, but 
there is no guarantee that no other cograph is closer to the perturbed 
graph. 
To simulate cographs based on their recursive construction definition we 
start with a graph on all vertices without any edges. 
We put all vertices in different bins and randomly merge 
bins. When two bins are merged, with some probability $d$ we connect all 
vertices which are in different bins.

\sloppy
We simulate cographs with different numbers of vertices
$n \in \{10,20,50,100\}$ and edge densities 
$d \in \{10\%,20\%,50\%\}$ where edge density is defined as the 
number of edges in a graph divided by the number of edges in a fully 
connected graph. 
We limit our evaluations to $d \leq 50\%$; edge densities of $x\%$ and 
$(100-x)\%$ will
produce the same results as the complement of a cograph is again a cograph.
As our simulation does not force the exact edge density, we 
exclude instances where the simulated cograph's edge density deviates by more 
than 10\% of the intended edge density.
For each parameter setting 100 cographs are computed: these are the \textit{true cographs}.
Each true cograph is then perturbed by randomly flipping vertex pairs---making 
edges non-edges and vice versa---to produce a \textit{noisy graph}, which will be given
as input to the heuristics. An edge change is only valid if it 
introduces at least one new $P_4$. Each edge can only be flipped once. We 
use noise rates $r \in \{1\%,5\%, 10\%, 20\%\}$. If it is not possible to 
introduce a new $P_4$ in each iteration step, we simulate and perturb a new 
cograph instead. It must be noted, that a flipped edge that creates a new
$P_4$ will be retained even if it also removes one or more existing $P_4$s
from the graph.

The heuristic solution to each noisy graph will be denoted the \textit{heuristic cograph}.
A noise rate of 1\% on graphs with 10 vertices is not interesting as
these graphs only contain 45 vertex pairs, so this parameter combination is 
excluded from evaluation.

The \emph{distance} between graphs
is the number of edge deletions and insertions needed to transform one graph 
into another.
Dividing this distance by the number of edges in a complete graph with the
same number of vertices gives the \emph{normalized distance}, a value between 0 and 1,
inclusive.
In the context of an instance of the \textsc{Cograph Editing} problem, the \emph{cost}
of a graph is simply the distance between it and the
input graph; we use solution cost as a measure of solution quality.
Given two pairs of graphs, the first pair is \textit{closer} than the 
second pair if the distance between the first pair is lower than
that between the second pair.

We do two kinds of evaluation: the first measures the quality of our
heuristics in solving the \textsc{Cograph Editing} problem, while the second
gauges the strength of the cograph property and this problem formulation to 
recover information from noisy data.

First, to test whether the heuristics produce 
good results we count how often a heuristic can find a cograph of cost
less than or equal to the cost of the true cograph. Recall that the true cograph gives the best upper bound for 
the \textsc{Cograph Editing} problem we can get in practice. Such a solution 
will be denoted ``fit''. It is clear, that this is 
not necessarily the optimal solution.

Second, we evaluate whether the heuristic solution ``improves on'' the noisy
input graph: that is, whether it produces a cograph which 
is closer to the true cograph than the noisy graph is to the true cograph. 
In applications like phylogenetic tree estimation, recovering the
structure of the underlying true cograph is of much more interest 
than a minimum number of modifications.
The use of this optimization problem (and our heuristic as approximation) is 
only justified to the
extent that a cograph that requires few edits usually corresponds closely
to this ``ground truth'' cograph.
Let $d$ be the distance between the true cograph and the noisy graph obtained through experimental measurements.
If it is frequently the case that there exist multiple different cographs at the same distance $d$ from the noisy graph, some of which are at large distances from the true cograph, then even an exact solution to the optimization problem is of limited use in such applications.
Worse yet, if it is common to find such cographs at distances strictly below 
$d$, then such an optimization problem is positively misleading.

\begin{figure}[tbp]
	\centering
	\includegraphics[width=\linewidth]{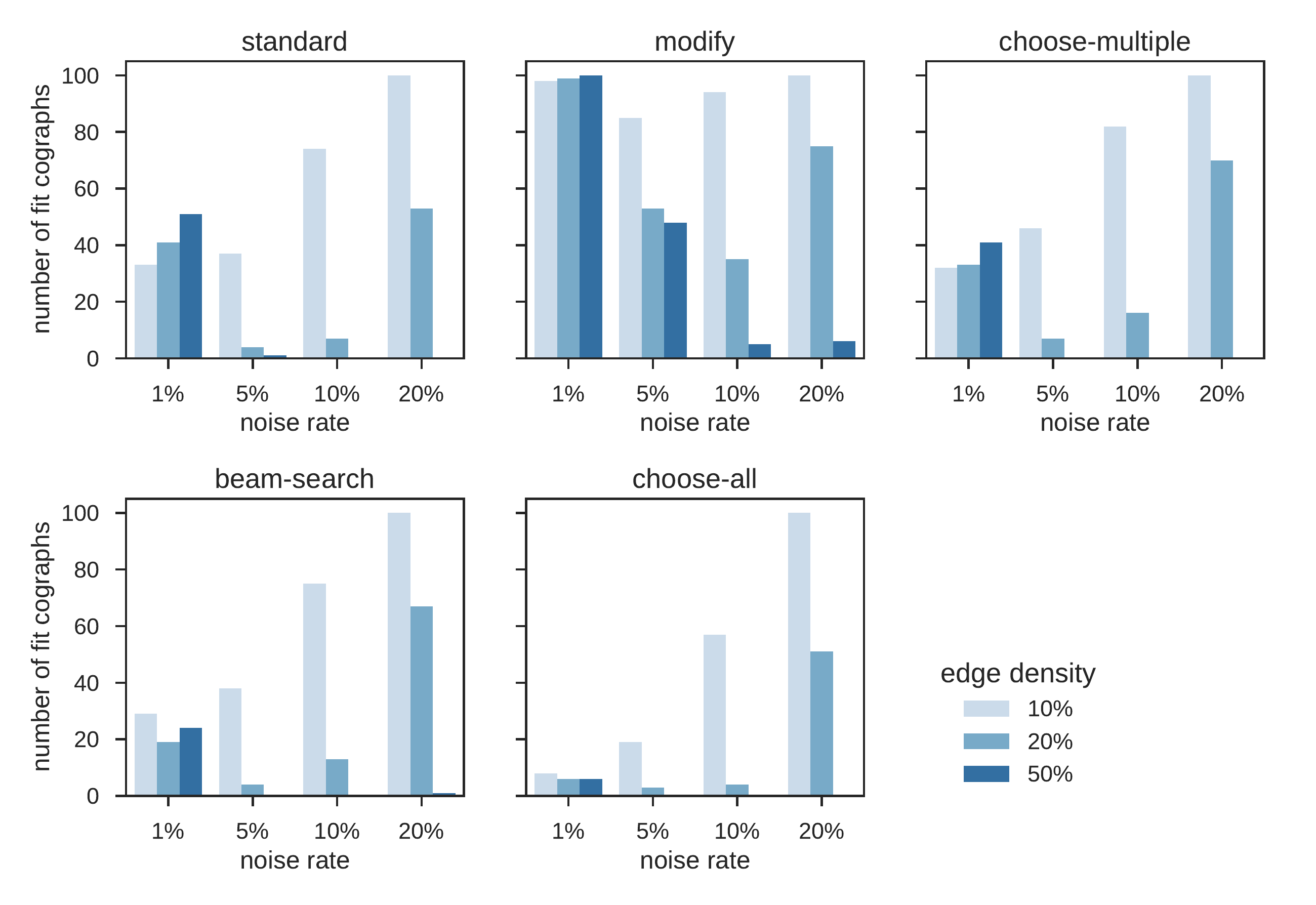}
	\caption{Comparison of five heuristic versions on graphs with 
		50 vertices. The figures show in how many cases the heuristics can 
		find a fit cograph---that is, a cograph that is at least as close to
		the noisy graph as the true cograph is to the noisy graph.}
	\label{fig:method_comp}
\end{figure}

On graphs of 20 vertices or fewer, all modifications perform quite well. 
To determine the best heuristic method on larger graphs, we compare results on graphs 
with 50 vertices (see Fig.~\ref{fig:method_comp}).
Here, the \textit{modify} heuristic clearly outperforms the other versions. Hence, 
it is interesting to see how this method performs on graphs with different numbers 
of vertices.

\begin{figure}[tbp]
	\centering
	\includegraphics[width=\linewidth]{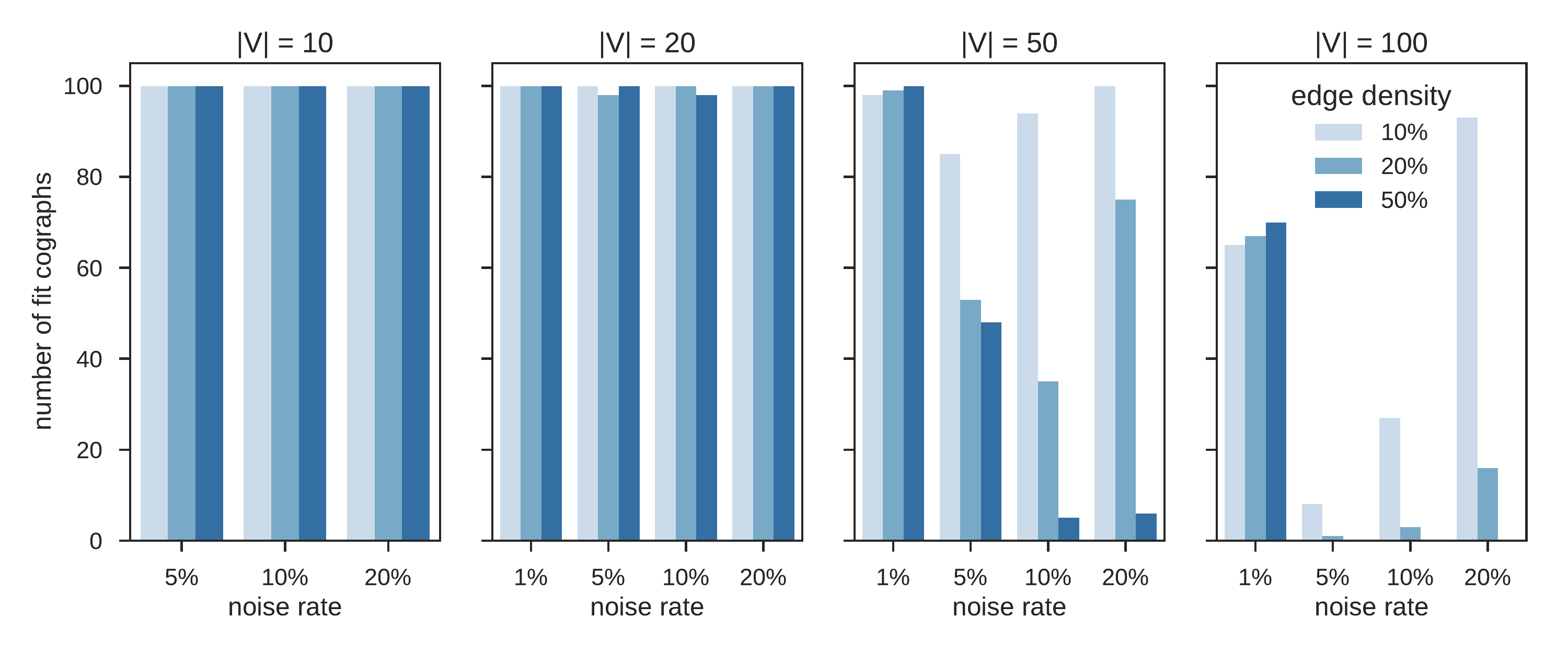}
	\caption{Performance of \textit{modify} heuristic. The figures show in how many cases the heuristics can 
	find a fit cograph---that is, a cograph that is at least as close to
	the noisy graph as the true cograph is to the noisy graph.}
	\label{fig:method_modify}
\end{figure}

For small graphs with 10 or 20 vertices the \textit{modify} heuristic finds 
a fit solution in almost all cases (see Fig.~\ref{fig:method_modify}),
as do the other heuristics. If 
input graphs have as little as 1\% noise, even on graphs with 50 vertices 
a fit 
cograph is found in over 98\%. For 100 vertices it is still over 65\%.
For more complex graphs, having a more balanced ratio of edges and 
non-edges, the number of fit solutions decreases. 
Interestingly, looking only at graphs with 100 vertices and over 1\% 
noise, high noise rates seem to favor a good heuristic solution. 
This is likely due to the fact that for high noise the true cograph 
is no longer a good bound on the optimal solution. 

\begin{figure}[tbp]
	\centering
	\includegraphics[width=\linewidth]{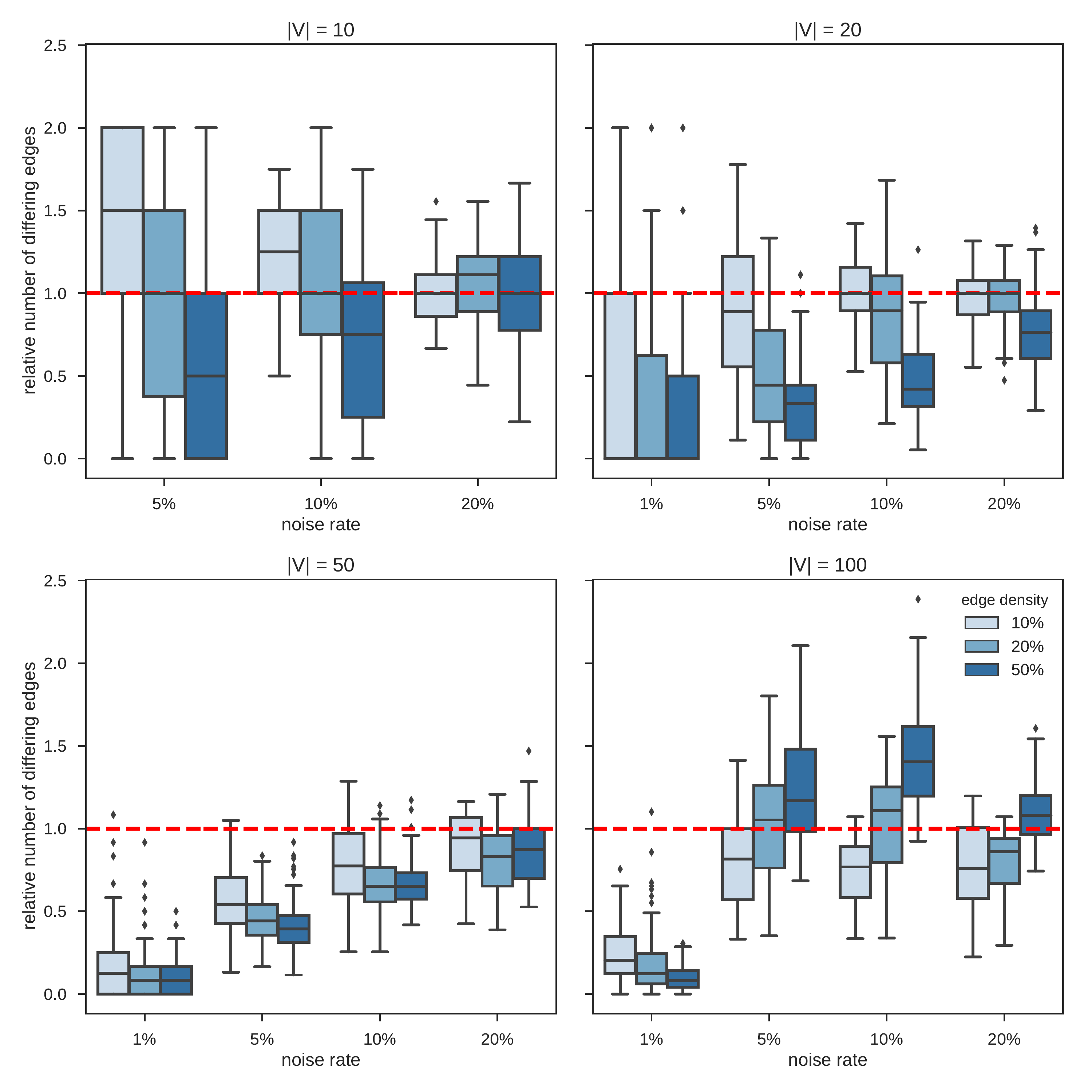}
	\caption{
		It is shown how well the \textit{modify} heuristic is able to 
		recover the ground truth.
		Each column summarizes the
		distribution of relative distances for 100 simulated instances.		
		A value lower than one indicates improvement; zero corresponds to perfect 
		recovery.  
	}
	\label{fig:similarity_modify}
\end{figure}

\begin{figure}[tbp]
	\centering
	\includegraphics[width=\linewidth]{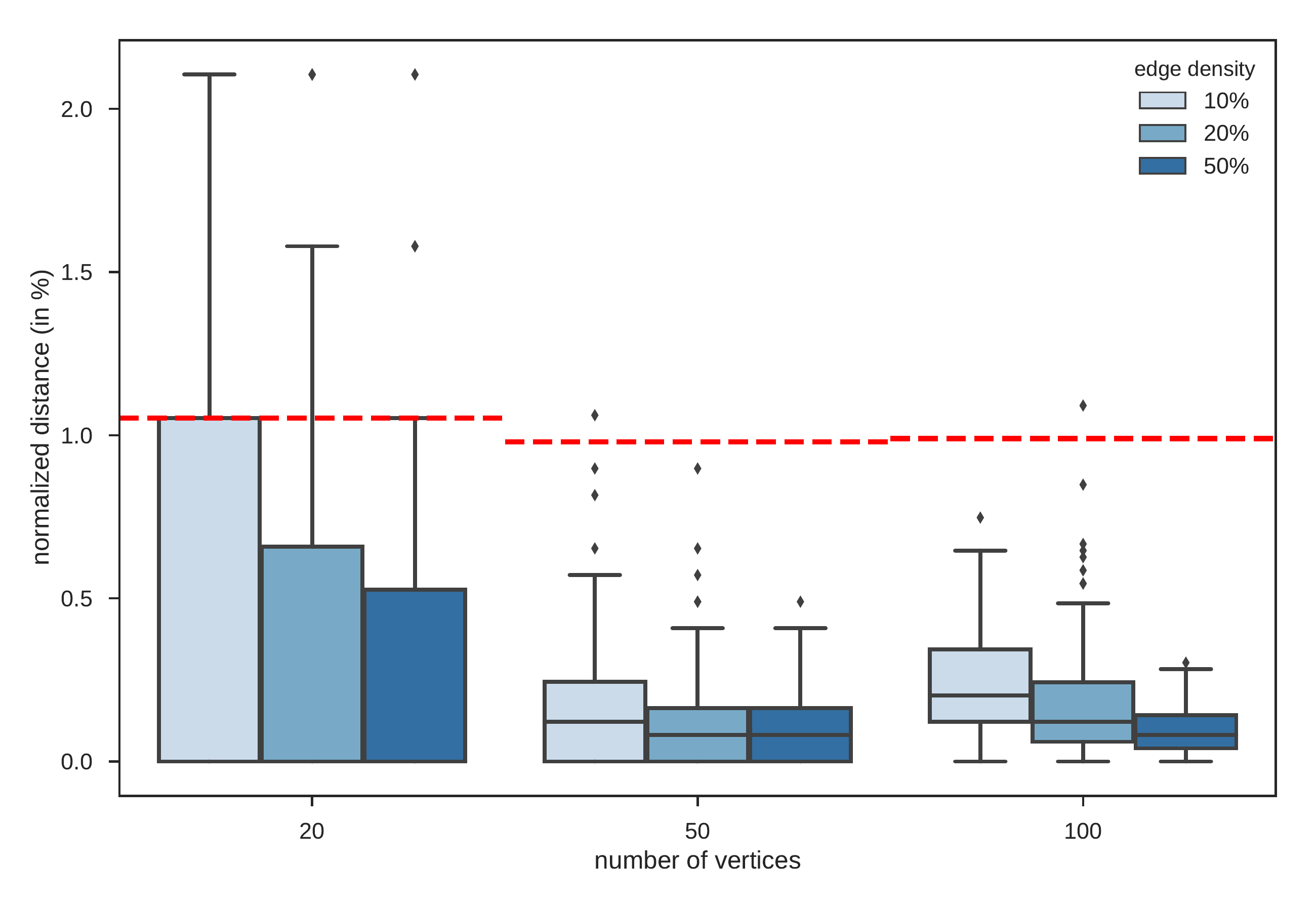}
	\caption{
		It is shown how well the \textit{modify} heuristic is able to 
		recover the ground truth for a intended noise rate of 1\%. Each 
		column summarizes the 
		distribution of normalized distances between a true cograph and the
		cograph computed heuristically from a noisy version of it for 100 
		simulated instances. The red dashed line indicates the normalized 
		distance between the noisy graph and the true cograph (noise rate). 
		There are deviations from the intended noise rate as the number of 
		edges in a graph is not continuous. Values below this line show 
		improvement.
	}
	\label{suppl:fig:similarity_modify_1noise}
\end{figure}

In application the relevant question is whether or how well the true cograph 
can be recovered from noisy data. To make different parameter combinations 
comparable we evaluate relative distances. Given distances 
$dist_{n}$ between the true cograph and the noisy graph and $dist_{h}$ between the true 
and the heuristic cographs, the \emph{relative distance} is $dist_{rel} = 
\frac{dist_{h}}{dist_{n}}$ (see Fig.~\ref{fig:similarity_modify}). A value 
smaller than one implies an improvement: the true cograph is closer to 
the heuristic cograph than to the noisy graph.  A $dist_{rel}$ larger than 
one implies a loss of 
similarity, while a value of zero corresponds to a perfect match 
between heuristic and true cograph.
The median $dist_{rel}$ for graphs with 20 vertices and 1\% noise is 0.0\,. 
The 
mean is 0.54, 0.34 and 0.27 for 10\%, 20\% and 50\% edge density, 
respectively.
The distances relative to the maximum number of possible edges can 
be seen in supplementary Fig.~\ref{suppl:fig:similarity_modify_rest} and 
\ref{suppl:fig:similarity_modify_1noise}.

Interestingly, for graphs of size 10 to 50, a certain amount of complexity, 
meaning greater edge density, seems to encourage a better recovery of the 
true cograph. This might be due to the fact that on sparse graphs there are 
often multiple options to resolve a $P_4$ which all lead to good results. 
Hence, there is no unambiguous way to denoise the graph.
Particularly on graphs with 50 vertices we see that increasing edge density 
leads to fewer fit cograph solutions (see Fig.~\ref{fig:method_modify}), 
but on average the resulting graph is closer to the true cograph (see Fig.~\ref{fig:similarity_modify}). 
This observation does not hold for graphs with 100 vertices;
but, as already explained, the true cograph no longer gives a good cost bound for large noisy 
graphs and so we also cannot expect to recover it.
If we limit our evaluation to graphs with 10 and 20 vertices, we are able to
find a fit cograph in almost all cases. 
The complexity of graphs with 10 vertices seems to be not sufficiently high 
to reliably produce a cograph closer to the true cograph.
For graphs of size 20, heuristic and true cograph are mostly closer to each 
other than the noisy graph is to the true graph. This means we are able to 
partially recover the ground truth.
Nevertheless, only for 1\% noise and at least 20 vertices can we either 
recover the true cograph or at least get very close to the correct solution.

\begin{figure}[tbp]
	\centering
	\includegraphics[width=0.8\linewidth]{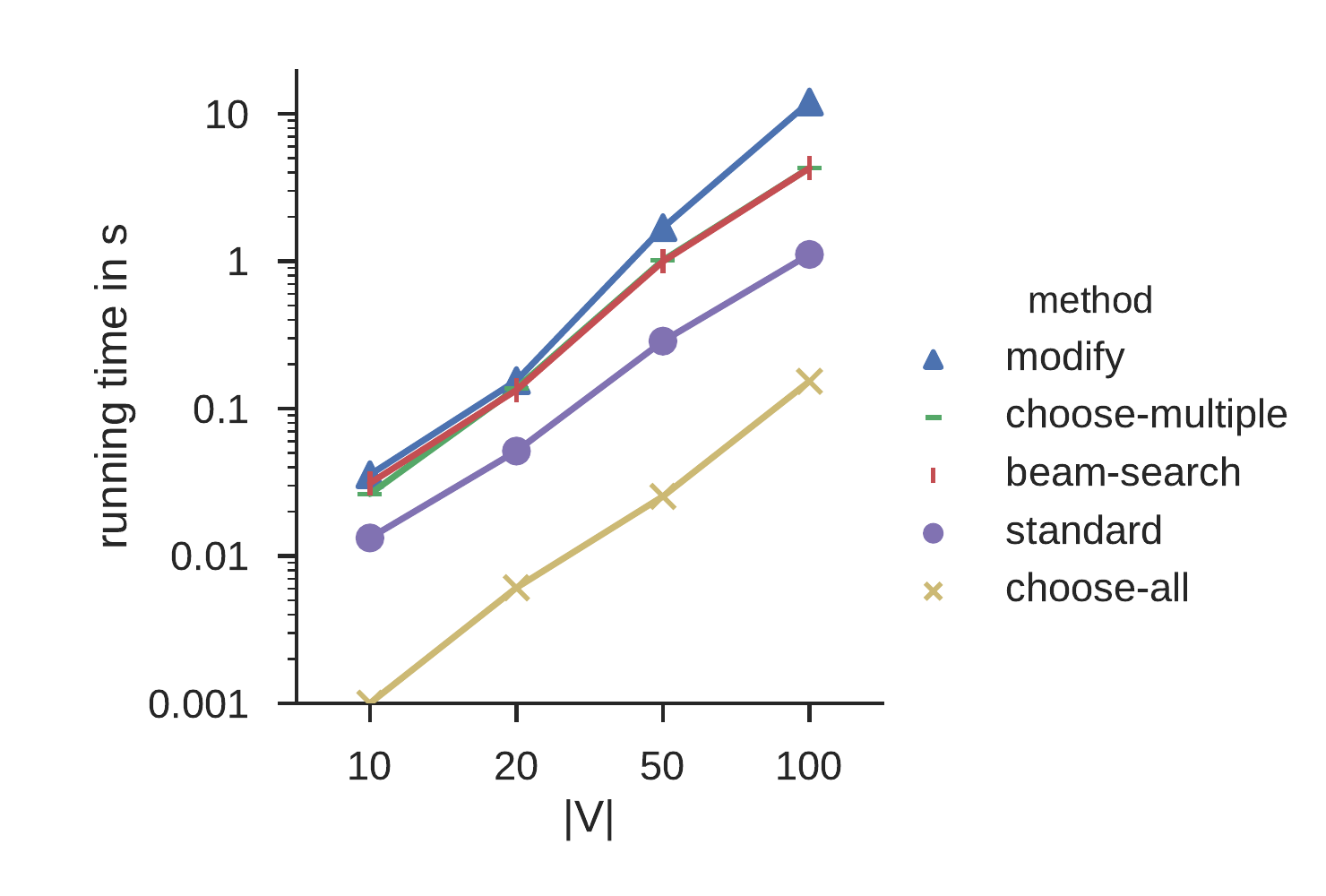}
	\caption{Running times for five heuristic versions. Noise 
		rate is 5\%, edge density is 20\%. Remaining parameter 
		combinations show similar results. The methods 
		\textit{choose-multiple} and \textit{beam-search} have almost 
		identical running times.}
	\label{fig:runningtimes}
\end{figure}

Lokshtanov's algorithm has a running time linear in the number of vertices 
plus 
edges. The \textit{standard} heuristic is just the second fastest method in 
our evaluation because we run it 100 times and \textit{choose-all} only once 
(see Fig.~\ref{fig:runningtimes}). As expected, running times of 
\textit{beam-search} and \textit{choose-multiple} are both slower than 
\textit{standard}. 
An iteration step in Lokshtanov's algorithm for adding $v_i$ to $H_i$ is 
composed of two actions. Step A consists of examining which edges need to be 
added so that $H_i$ is a minimal cograph completion of $G_i$. In step B, 
$v_i$ and all necessary edges are added to $H_i$  
(more precisely, to its cotree). 
Both \textit{choose-multiple} and \textit{beam-search} perform step A 
ten times more often than \textit{standard}, but all three methods perform 
the same number of B-steps.
Hence, running times do not increase by a factor of ten but rather by two to 
four.

On graphs with 50 vertices no method takes more than 1.69 seconds on 
average; 
For 100 vertices the slowest method is \textit{modify} with 12.03 seconds on 
average. Running times of \textit{modify} grow fastest. Still, it is easily 
applicable to graphs with several hundreds of vertices.
The \textit{choose-all} modification is fastest because only a single 
cograph is computed; but running times grow faster than for 
\textit{standard}, 
\textit{choose-multiple} and \textit{beam-search}.
All computations were executed single-threaded on an Intel E5-2630 @ 2.3GHz.

\section{Conclusion}

We presented an exact algorithm solving the weighted \textsc{Cograph 
Editing} problem in $O(3^n n)$ time and $O(2^n)$ space.
We evaluated five heuristics based on an algorithm for minimal cograph 
completions.
For small and medium graphs of 10 and 20 vertices we are able to find 
cographs with equal or lower cost than the ground truth, indicating 
that we find (nearly) optimal solutions. 
In application, the focus lies on recovering the true cograph, not the 
optimal one. We showed that for small noise of 1\% we get results very 
similar to this true cograph, even for large graphs with 100 vertices. 
Interestingly, it is easier to recover the true edges when graphs contain 
about 50\% edges.
For higher noise rates it is not possible to recover the true cograph. This 
may be partly explained by the fact that we apply a heuristic and do not solve the  
\textsc{Cograph Completion} problem optimally. But this observation already 
holds for medium graphs with 20 vertices on which we produce good 
results. We therefore argue that the cograph 
constraint is not strict enough to always correctly resolve graphs with 5\% 
noise and more. Therefore, if true graph structure recovery is important, 
low noise rates are crucial.

The presented heuristics are fast enough to be applied to graphs with 
several hundreds of vertices. 
Accuracy clearly improves when removing the restriction that in each 
iteration step edges can only be added or deleted.
Different heuristic modifications can be easily combined. This will likely 
improve results further.

\section*{Acknowledgment}
Funding was provided to W. Timothy J. White and Marcus Ludwig by Deutsche 
Forschungsgemeinschaft (grant BO~1910/9).

\bibliography{references}

\bibliographystyle{ieeetr}

%\appendix
\beginsupplement
\newpage
\FloatBarrier
\section{Supplementary}
\begin{figure}[tbp]
	\centering
	\includegraphics[width=\linewidth]{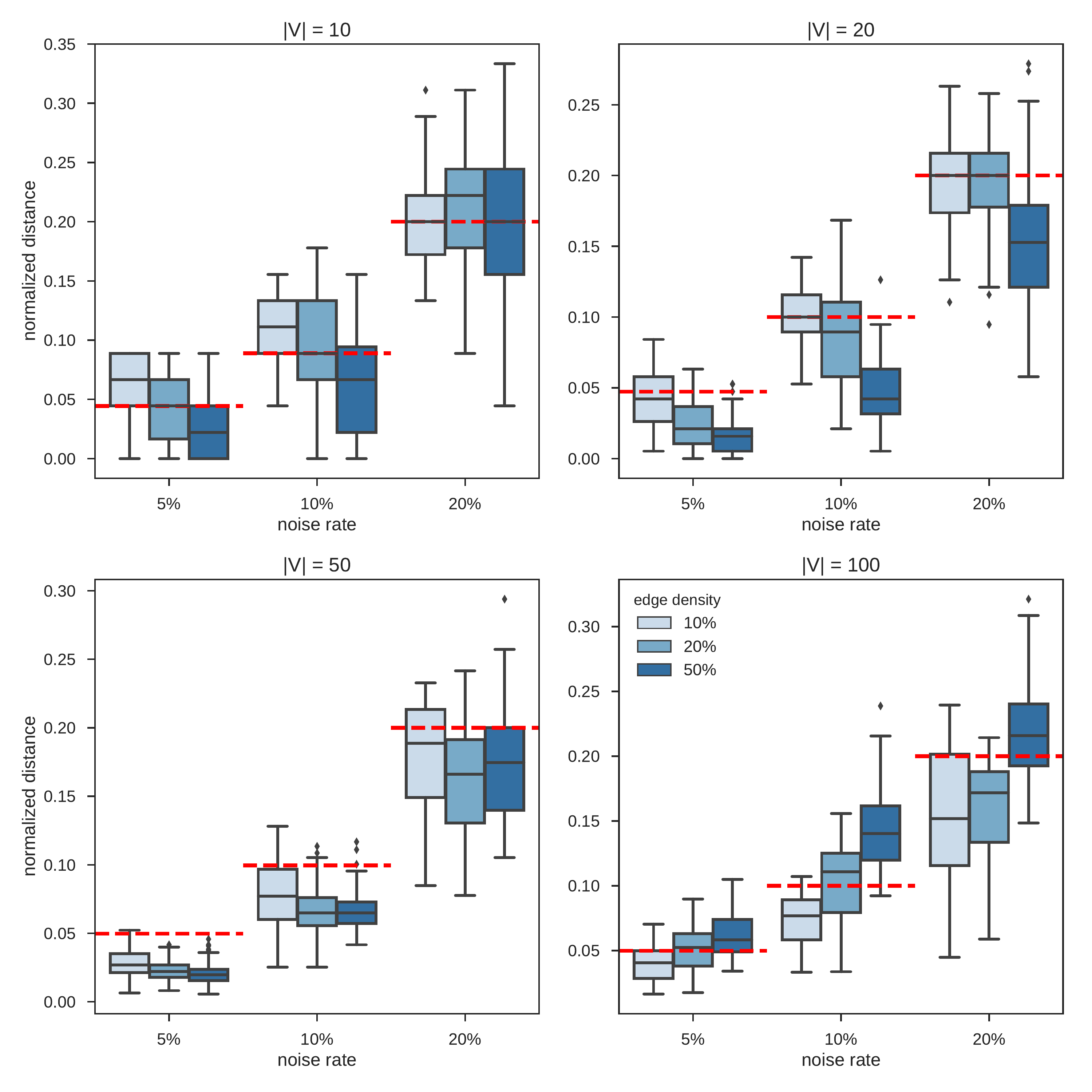}
	\caption{
		It is shown how well the \textit{modify} heuristic is able to 
		recover the ground truth for noise rates 5\%, 10\% and 20\%. Each 
		column summarizes the
		distribution of normalized distances between a true cograph and the
		cograph computed heuristically from a noisy version of it for 100 
		simulated instances. The red dashed line indicates the normalized 
		distance between the noisy graph and the true cograph (noise rate). 
		There are minor deviations from the intended noise rate as the 
		number of 
		edges in a graph is not continuous. Values below this line show 
		improvement.
	}
	\label{suppl:fig:similarity_modify_rest}
\end{figure}

\end{document}